\documentclass[prd,showpacs,aps,preprint]{revtex4}
\usepackage{graphicx}
\usepackage{dcolumn}
\usepackage{amsfonts}
\usepackage{amsmath}
\def \be{\begin{equation}}
\def \ee{\end{equation}}
\def \bea{\begin{eqnarray}}
\def \eea{\end{eqnarray}}
\def \p{\partial}
\def \nn{\nonumber}

\newcommand{\ep}{\epsilon}

\newcommand{\lraw}{\longrightarrow}

\newcommand{\td}{\tilde}

\newcommand{\beq}[1]{\begin{eqnarray}\label{#1}}
\newcommand{\eeq}{\end{eqnarray}}

\newcommand{\llraw}{-\!-\!\!\!\lraw}
\newcommand{\CN}[1]{{\cal N}=#1}
\newcommand{\bint}{-\!\!\!\!\!\!\int}
\begin{document}
\preprint{USTC-ICTS-04-06}
\title{Open Spin Chain and Open Spinning String}
\vspace{0.2in}
\author{Bin Chen$^1$}
\email{bchen@itp.ac.cn}
\author{Xiao-Jun Wang$^2$}
\email{wangxj@ustc.edu.cn}
\author{Yong-Shi Wu$^3$}
\email{wu@physics.utah.edu} \affiliation{ \centerline{$^1$
Interdisciplinary Center for Theoretical Study}
\centerline{Chinese Academy of Sciences, Beijing 100080, China}
\centerline{$^2$ Interdisciplinary Center for Theoretical Study}
\centerline{University of Science and Technology of China}
\centerline{An-Hui, He-Fei 230026, China} \centerline{$^3$
Department of Physics, University of Utah} \centerline{Salt Lake
City, Utah 84112, USA}}
\begin{abstract}
In this letter, we study the open spinning strings and their SYM
duals. A new class of folded open spinning strings is found. At
planar one-loop level in SYM, by solving the thermodynamic limit
of the Bethe ansatz equations for an integrable open spin chain,
we find good agreement with string theory predictions for energies
of both circular and folded two-spin solutions. A universal
relation between the open and closed spinning strings is verified
in the spin chain approach.
\end{abstract}
 \maketitle

\section{Introduction}
Recently the integrable spin chains have become a powerful tool to
study string/gauge theory duality \cite{Mald1}. On one hand, two
special limits of IIB string theory in $AdS_5\times S^5$ have been
identified, the plane-wave limit \cite{BFHP,BMN02} and the
semi-classical limit \cite{GKP02}, in either of which the free
closed string spectrum is known \cite{Mets02,FT}. On the other
hand, on the dual gauge theory ($D=4$ $\CN{4}$ super Yang-Mills)
side, to test these string predictions, one needs to compute the
anomalous dimensions \cite{GKP98,Witten98} of corresponding gauge
invariant composite operators. (For a recent review, see
\cite{SS03,Tsey03} and references in.) In either of the
above-mentioned limits, one needs to deal with mixing of a huge
number of very long composite operators.

It was first observed in \cite{MZ02} that at least at the planar
one-loop level, the mixing matrix of composite operators in a
certain sector is equivalent to the Hamiltonian of an integrable
spin chain. Thus the difficult problem of diagonalizing the
anomalous dimension matrix of a huge number of long composite
operators is reduced to solving the spin chain in a systematic
manner with the algebraic Bethe Ansatz equations (ABAE)
\cite{ABAE}. The great advantage of this approach is that by
considering the thermodynamic limit of the Bethe ansatz, in the
spirit of ref. \cite{YY69}, it can easily go beyond the BMN regime
and be extended to the cases with many ``impurities'', such as the
semi-classical spinning strings \cite{BMSZ03,KMMZ04}. Also it is
accessible by numerical methods for both plane-wave strings and
spinning strings as well \cite{BFST,Beisert03}. (For integrable
spin chains in certain sectors of orbifold gauge theories with
less supersymmetries (${\CN 1,2}$), see \cite{WW03}.)

Theory with both open and closed strings is known to have richer
physics. The plane-wave/SYM duality has also been generalized to
incorporate open strings \cite{BGMNN02,Imamura}. It is shown
recently \cite{CWW03} that both Neumann and Dirichlet boundary
conditions for open strings correspond, in the dual gauge theory,
to integrable boundary terms in the pertinent open spin chain
Hamiltonian. (In \cite{Wolfe04}, the open integrable system with
boundary has been identified in another context.) So the ABAE
approach, together with whole open spin chain machinery (see,
e.g., \cite{Skly88,BYB,VG94,DN98}), can also be applied to the
gauge theory that is dual to an open+closed string theory. This
result is of principal interests, in that it extends the
integrability to open strings and their gauge duals as well. Thus
the appearance of integrable spin chain seems not accidental at
all. To provide more evidence for the last statement, in this
letter we use the integrable open spin chain approach to study the
newly found open spinning strings \cite{Stef03b}, and compare the
results from the ABAE with the string predictions. As will be
seen, once again we find good agreements. In this way, we add more
evidence to the belief that there must be a profound relationship
between string theory and spin chains.

In Sec. 2 we present folded open spinning strings, including a new
class of solutions that rotate both in $AdS_5$ and $S^5$. Their
energies is reproduced at planar one-loop level in SYM by solving
the ABAE for an integrable open spin chain. In Sec. 4 we use the
same approach to study the SYM dual of circular open spinning
strings and of its fluctuations and get good agreement too. In
both cases, an universal relation between closed and open spinning
strings is verified in the spin chain approach. Sec. 5 is devoted
to a short summary.

\section{Folded Open Spinning Strings}

Here we follow the notations in \cite{Tsey03}, and present some
open spinning string solutions that are relevant to our AdS/CFT
test in the subsequent section. We will consider strings in
$AdS_5\times S^5/Z_2$ orientifold, whose dual is a conformal ${\CN
2}$ $Sp(N/2)$ gauge theory with matter \cite{BGMNN02}. For
simplicity, we will suppress Chan-Paton factors, which are not
essential to our discussions.

The bosonic part of the action for a string in $AdS_5 \times S^5$
reads
 \beq{1}
 I=\frac{\sqrt{\lambda}}{2\pi}\int d\tau d\sigma(L_{AdS}+L_S)
 \eeq
 where
 \beq{2}
 L_{AdS}&=&-\frac{1}{2}\eta^{PQ}\p_aY_P\p^aY_Q
 +\frac{1}{2}\tilde{\Lambda}(\eta^{PQ}Y_PY_Q+1),\hspace{0.7in}
 P,Q=0,\cdots ,5,\nn \\
 L_{S}&=&-\frac{1}{2}g^{MN}\p_aX_M\p^aX_N+\frac{1}{2}
 {\Lambda}(g^{MN}X_MX_N-1),\hspace{0.35in} M,N=1,\cdots ,6,
 \eeq
with the metric $g^{MN}=\delta^{MN}$ and $\eta^{PQ}={\rm
diag}\{-1,1,1,1,1,-1\}$. Here $\Lambda,\;\tilde{\Lambda}$ are the
Lagrange multipliers imposing the constraints, respectively,
 \beq{3}
 \eta^{PQ}Y_PY_Q=-1, \hspace{5ex} g^{MN}X_MX_N=1
 \eeq

As usual, for a closed string, the range of $\sigma$ is from $0$
to $2\pi$, and $X_M, Y_P$ satisfy periodic boundary conditions.
For an open string, $\sigma$ ranges from $0$ to $\pi$ and $X_M,
Y_P$ satisfy the Neumann or Dirichlet boundary conditions. It is
often convenient to use the global coordinates:
 \beq{4}
 Y_1+iY_2&=&\sinh\rho\sin\theta e^{i\phi_1},
 \quad\quad\quad Z'=X_1+iX_2=\sin\gamma\cos\psi e^{i\varphi_1} \nn \\
 Y_3+iY_4&=&\sinh\rho\cos\theta e^{i\phi_2},
 \quad\quad\quad W=X_3+iX_4=\sin\gamma\sin\psi e^{i\varphi_2}\nn \\
 Y_5+iY_0&=&\cosh\rho e^{i\phi_3},
 \quad\quad\quad\quad\quad Z=X_5+iX_6=\cos\gamma e^{i\varphi_3}
 \eeq

There are two classes of classical spinning string solutions:
circular and folded. We first consider the folded ones, since
their treatment in the spin-chain approach is easier. The folded
closed two-spin classical solutions have been constructed in
\cite{FT}, and applied to test AdS/CFT far beyond the BPS regime
\cite{BFST}. There are two types of two-spin strings: those
rotating on both $AdS_5$ and $S^5$, denoted as $(S,J)$, and those
rotating only on $S^5$, denoted as $(J_1,J_2)$. They are actually
related to each other by analytical continuation. On the gauge
theory side, the corresponding gauge invariant operators are,
respectively,
 \beq{11}
 Tr{\cal D}^SZ^J + \cdots, \hspace{5ex}
 TrZ^{J_1}\Phi^{J_2}+\cdots.
 \eeq

For an open string, one can construct similar classical solutions.
For the $(S,J)$-type, one has
 \beq{12}
 \theta=0,~~&\rho=\rho(\sigma),~~& \phi_3=k\tau,
 \nn \\
 \gamma=0,~~&\phi_2=\omega\tau, ~~& \varphi_3=\nu\tau
 \eeq
with $\rho$ satisfying the Neumann boundary condition $\rho^\prime
|_{\sigma=0,\pi}=0.$

From the equation of motion and Virasoro constraints, we have
 \beq{14}
 \rho^{\prime\prime}&=&\sinh\rho\cosh\rho(k^2-\omega^2) \nn \\
 \rho^{\prime 2}&=&k^2\cosh^2\rho-\omega^2\sinh^2\rho-\nu^2
 \eeq
In the closed string one-fold case, $\sigma$ ranges from $0$ to
$2\pi$ and could be divided into four segments: $\rho(\sigma)$
increases from $0$ to $\rho_0$ when $\sigma$ increases from $0$
(or $\pi$) to $\pi/2$ (or $3\pi/2)$, and then decreases back to
$0$ when $\sigma$ goes on to $\pi$ (or $2\pi$). Here $\rho_0$ is
the maximal value of $\rho$, determined by
 \beq{15}
 k^2\cosh^2\rho_0-\omega^2\sinh^2\rho_0-\nu^2=0.
 \eeq
Therefore the turning point is at $\sigma=\frac{\pi}{2}$ or
$\frac{3\pi}{2}$, and $\sigma=0,\pi$ are two singular points
(fold-point). However, one has the freedom to translate the above
points, keeping the relative distance unchanged. So we can move
the turning points to $\sigma=0,\pi$, where $\rho^\prime=0$, and
the singular point to $\sigma=\frac{\pi}{2}$. This is just what we
need for a folded open string to satisfy the Neumann boundary
condition. In this case, $\sigma$ ranges from $0$ to $\pi$ and
includes two segments. The energy, spin and angular momentum are
 \beq{16}
 E&=&2\frac{\sqrt{\lambda}}{2\pi}k\int^{\rho_0}_0d\rho
 \frac{\cosh^2\rho}{\sqrt{k^2\cosh^2\rho-\omega^2\sinh^2\rho-\nu^2}}
 \nn \\
 S&=&2\frac{\sqrt{\lambda}}{2\pi}\omega\int^{\rho_0}_0d\rho
 \frac{\sinh^2\rho}{\sqrt{k^2\cosh^2\rho-\omega^2\sinh^2\rho-\nu^2}}\\
 J&=&2\frac{\sqrt{\lambda}}{2\pi}\nu\int^{\rho_0}_0d\rho
 \frac{1}{\sqrt{k^2\cosh^2\rho-\omega^2\sinh^2\rho-\nu^2}}. \nn
 \eeq
The factor of $2$ indicates that we have two segments of the
string stretching from $0$ to $\rho_0$. This class of folded
spinning string solutions are a new result of our paper. Compared
with the closed string case, we have
 \beq{17}
 E_o=\frac{1}{2}E_c(2S,2J).
 \eeq
The dual operator in the gauge theory is proportional to
 \beq{18}
 Q\Omega {\cal D}^S(Z\Omega)^JQ+\cdots.
 \eeq

For the $(J_1,J_2)$-type solution~\cite{Stef03b}, we have the
ansatz
 \beq{19}
 \rho=0,~~&\phi_3=k\tau,~~&\gamma=\frac{\pi}{2} \nn\\
 \psi=\psi(\sigma),~~&\varphi_1=\omega_1\tau,~~&\varphi_2=\omega_2\tau.
 \eeq
The equations of motion and Virasoro constraints are
 \beq{20}
 \psi^{\prime\prime}+\frac{1}{2}\omega^2_{21}\sin2\psi&=&0 \nn \\
 \psi^{\prime 2}+\omega_1^2\cos^2\psi+\omega_2^2\sin^2\psi&=&k^2
 \eeq
where we have assumed without losing generality that
$\omega^2_{21}=\omega^2_{2}-\omega^2_{1}>0$. In the same spirit as
in above, one can put the turning point at $\sigma=0,\pi$ and
fold-point at $\sigma=\frac{\pi}{2}$. The maximal value of
$\psi_0$ is determined by
 \beq{21}
 \omega_1^2\cos^2\psi_0+\omega_2^2\sin^2\psi_0=k^2.
 \eeq
Also the relation between energy and angular momenta can be read
off from the closed string result:
 \beq{22}
 E_o(J_1,J_2)=\frac{1}{2}E_c(2J_1,2J_2).
 \eeq
The dual operators is of the form
 \beq{23}
 Q\Omega(Z\Omega)^{J_1}(Z'\Omega)^{J_2}Q+\cdots.
 \eeq

In the following, we will show that on the gauge theory side, the
anomalous dimensions of operators (\ref{23}) do respect the
relation (\ref{22}) between energy and angular momenta, verifying
the AdS/CFT correspondence for folded two-spin open strings.

\section{SYM Dual from Integrable Open Spin Chain}

It was noticed in \cite{CWW03} that an integrable structure exists
in the ${\CN 2}$ $Sp(N/2)$ gauge theory, at least at planar 1-loop
level after turning off string interactions. More precisely, the
anomalous dimension matrix for the open string BMN operators in
the holomorphic scalar sector can be identified with the
Hamiltonian of an $SU(3)$ open spin chain with integrable boundary
terms. This observation provides a framework for dealing with
one-loop mixing of a huge number of operators in SYM. We expect to
recover the spectrum of classical spinning string
solution~(\ref{22}) by solving the integrable spin chain.

For an integrable $SU(n)$ open spin chain, the Hamiltonian is
given by \cite{Skly88,DN98}
\beq{24}H_{open}=\sum_{m=1}^{L-1}H_{m,m+1}
+\frac{1}{2\xi_-}\left.\frac{d}{du}
K^-_{1,(l)}(u,\xi_-)\right|_{u=0} +\frac{{\rm
tr}_0[K^+_{0,(l)}(0,\xi_+)H_{L,0}]} {{\rm tr}K^+_{(l)}(0,\xi_+)},
\eeq where $H_{m,m+1}=I_{m,m+1}-P_{m,m+1}$ with $P_{m,m+1}$ the
permutation operator. $K_{i,(l)}$ denotes the $K$-matrix acting on
the Hilbert space ${\cal H}_i$ associated with the $i$-th site
(${\cal H}_0$ being the auxiliary space). The general diagonal
$K$-matrix ensuring integrability has been obtained in
\cite{VG94}: \beq{25}K^{\pm}_{(l)}(u,\xi_{\pm})={\rm diag}
\{\overbrace{a^\pm,...,a^\pm}^{l},
\overbrace{b^\pm,...,b^\pm}^{n-l}\}, \eeq where
\beq{26}a^+&=&i(\xi_+-n)-u,\hspace{0.55in}b^+=i\xi_++u,
\nonumber\\
a^-&=&i\xi_-+u,\hspace{1in}b^-=i\xi_--u,
\eeq
with arbitrary $\xi_\pm$ and any $l\in\{1,...,n-1\}$. The one-loop
anomalous dimensions of open BMN operators consisting of
holomorphic scalars (or eigenvalues of the
Hamiltonian~(\ref{24}))$^1$\footnotetext[1]{The expression of
eigen-energies and the ABAE depend on the choice of the
pseudo-vacuum $\omega$. Here we choose
$\omega=(\vec{v}\otimes)^{L-1}\vec{v}$ with
$\vec{v}=(1,0,...,0)^T$.} are given by \cite{VG94}
\beq{27}
\gamma=\frac{\lambda}{4\pi^2}\sum_{j=1}^{n_1}\ep(\mu_{1,j}),
\hspace{0.5in}\ep(\mu)=\frac{4}{\mu^2+1}.
\eeq
Here $\mu_{i,j}$ (the Bethe roots) satisfy the ABAE
\beq{28}1&=&[e_{l-2\xi_-}(\mu_{l,k})e_{2\xi_+-l}(\mu_{l,k})\delta_{l,q}
+(1-\delta_{l,q})]\prod_{j=1}^{M_{q-1}}e_{-1}(\mu_{q,k}-\mu_{q-1,j})
e_{-1}(\mu_{q,k}+\mu_{q-1,j}) \nonumber \\
&&\times \prod_{j=1,\atop j\neq k}^{M_q}e_{2}(\mu_{q,k}-\mu_{q,j})
e_{2}(\mu_{q,k}+\mu_{q,j})
\prod_{j=1}^{M_{q+1}}e_{-1}(\mu_{q,k}-\mu_{q+1,j})
e_{-1}(\mu_{q,k}+\mu_{q+1,j}) \\
&& \mbox{for $k=1,\cdots, M_q$ and $q=1,\cdots, n-1$,} \nonumber
\eeq
with $M_0=L,\;M_n=0,\;\mu_{0,j}=\mu_{n,j}=0$, and
\beq{29}
e_n(\mu)=\frac{\mu+in}{\mu-in}. \eeq

For the SYM dual of folded open two-spin solutions, two boundary
terms in the Hamiltonian~(\ref{24}) should be of the form
$\Sigma_1=\Sigma(\otimes I_{3\times 3}
)^{L-1},\;\Sigma_L=(I_{3\times 3}\otimes)^{L-1}\Sigma$, with
$\Sigma={\rm diag}\{0,0,1\}$. The nonzero element in $\Sigma$
corresponds to the Dirichlet boundary condition \cite{BGMNN02},
while the zero diagonal elements in $\Sigma$ correspond to the
Neumann boundary condition. These requirements are satisfied by
taking $\xi_+=\xi_-=1$ after setting $n=3,\;l=2$ in the
$K$-matrix~(\ref{25}). The boundary parameters $\xi_{\pm}$ break
the bulk $SU(3)$ symmetry of the spin chain down to $SU(2)\times
U(1)$, the same as the $R$-symmetry of the gauge theory at hand.

For a folded open string of type $(J_1,J_2)$, the dual operator
(\ref{23}) consists of $J_2$ $Z^\prime$s inserted in $J_1$ $Z$s.
Or in open spin chain language, it consists of $L=J_1+J_2$ sites
and $J_2$ $\mu_{1,j}$-impurities, which are determined by the ABAE
 \beq{30}
 \left(\frac{\mu_j+i}{\mu_j-i}\right)^{2L} \prod^{J_2}_{\stackrel{k=1}{k\neq
 j}}\frac{\mu_j-\mu_k+2i}{\mu_j-\mu_k-2i} \cdot
  \frac{\mu_j+\mu_k+2i}{\mu_j+\mu_k-2i}.
 \eeq
Here we have used $\mu_j=\mu_{1,j}$ to simplify the notation. We
observe that these equations are invariant under $\mu_j \to
-\mu_j$ for any single $j$. This manifests the fact that for an
open chain $\mu_j$ labels a standing wave, so the sign of $\mu_j$
is insignificant. This motivates us to use the following trick to
solve these equations. Recall that the ABAE for a state of the
corresponding closed spin chain, labelled as $(2L, 2J_2)$, with
$2L$ the length of the spin chain and $2J_2$ the number of the
impurities, is
of the form 
 \beq{new}
 \left(\frac{\nu_j+i/2}{\nu_j-i/2} \right)^{2L} \prod^{2J_2}_{\stackrel{k=1}{k\neq
 j}}\frac{\nu_j-\nu_k+i}{\nu_j-\nu_k-i}\; .
 \eeq
When one requires that the Bethe roots distribute symmetrically,
which is the simplest assumption to satisfy the trace condition,
the above equations reduce to eq. (23).
In other words, a state of the open spin chain, labelled by
$(L,J_2)$, is related to the state of a closed spin labelled by
$(2L, 2J_2)$ with symmetrically distributed Bethe
roots$^2$\footnotetext[2]{For large $L$ one can ignore the minor
difference between the two ABAE's caused by the root
$\nu_k=-\nu_{j}$.}.

Since $\mu_j$ is of order $L$ for large $L$, one can rescale
$\mu$'s by $\mu_j= 4Lx_j$ and take the logarithm of the
ABAE~(\ref{30}). One has
 \beq{31}
 \frac{1}{x_j}=2\pi n_j+\frac{1}{L}\sum_{\stackrel{k=1}{k\neq
 j}}^{J_2}(\frac{1}{x_j-x_k}+\frac{1}{x_j+x_k}).
 \eeq
The above equation is the same as the case for a closed chain
state $(2L, 2J_2)$, Eq.(2.7) in \cite{BMSZ03}, with symmetrically
distributed Bethe roots.
Thus in the spin chain approach we have
 \beq{32}
 E_o=\frac{1}{2}E_c(2J_1,2J_2).
 \eeq
This relation is the same as that between the folded two-spin open
and closed strings suggested in ref. \cite{Stef03b}. In this way,
we have shown that the open spin chain in the gauge theory at
planar one-loop level reproduces the energies of a folded open
string. The $(S,J)$-type open spinning string we found in Sec. 2
should be described by an $SU(2,2|2)$ super open spin chain, for
which the relation (\ref{32}) is expected to be valid too.

What is essential to our above treatment is the following
observation for known folded $(J_1,J_2)$-type closed string
solutions: In the ``thermodynamic limit'' $L\to\infty$, the Bethe
roots in the complex plane are located along two cuts, which are
almost parallel to the imaginary axis and symmetrical under
reflection about the imaginary axis~\cite{BMSZ03}. In the above we
have been able to show that in the spin chain approach, in the
limit $L\to \infty$, the Bethe roots on one of the cuts can be
identified as the Bethe roots associated with an open spin chain
state. So the relation (\ref{32}) follows naturally. Since the
folded $(S,J)$-type closed spinning string is known to correspond
to a distribution of Bethe roots along two cuts located on the
real axis symmetrically about $\mu=0$~\cite{BFST}, we expect the
above trick should also work too.

\section{Circular Open Spinning String and Open Spin Chain }

Let us consider the circular two-spin open string solutions on
$R_t\times S^5$, which have been studied in~\cite{Stef03b}, with
the ansatz
 \beq{5}
 \rho=0, ~~& \phi_3=k\tau,~~ &\gamma=\gamma_0, \nn \\
 \psi=m\sigma,~~&
 \varphi_1=\varphi_2=\omega \tau, ~~&\varphi_3=\nu\tau,
 \eeq
where
 \beq{6}
 k^2&=& \nu^2+(\omega^2+m^2-\nu^2)\sin^2\gamma_0, \nn \\
 \omega^2&=&m^2+\nu^2.
 \eeq
The corresponding energy and spins are
 \beq{7}
 E&=&\frac{\sqrt{\lambda}}{2}k\; ,  \hspace{0.5in}
 J_3=\frac{\sqrt{\lambda}}{2}\nu[1-\frac{1}{2m^2}(k^2-\nu^2)],
 \nn \\
 J_1&=&J_2=\frac{\sqrt{\lambda}}{8m}\sqrt{1+(\frac{\nu}{m})^2}
 (k^2-\nu^2).
 \eeq

The open BMN string case corresponds to $\gamma_0=0$, so that
$k=\nu$, $J_1=J_2=0$ and $E=J_3=J$. The characteristic frequency
of the fluctuations around the classical solution,
$\omega_n=\pm\nu\pm\sqrt{n^2+\nu^2}$, leads to (for lower-energy
modes)
  \beq{9}
  \Delta E_o =-1+\sqrt{1+\frac{\lambda n^2}{4J^2}}=\Delta E_c(2J).
  \eeq
This is exactly the open string excited BMN spectrum in the
plane-wave background \cite{BGMNN02}, which can be recovered in
dual gauge theory at planar one-loop level from an integrable
$SU(3)$ open spin chain, as shown in \cite{CWW03}.

Another special case is $\gamma_0=\frac{\pi}{2}$, which leads to
$J_1=J_2=J, J_3=0$ and
 \beq{10}
 E=L\sqrt{1+\frac{m^2\lambda}{4L^2}}
 = \frac{1}{2}E_{c}(2J_1,2J_2),
 \eeq
where $L=J_1+J_2=2J$ and $E_c$ is the energy of corresponding
circular closed string. In this case, $Z'$ directions satisfy the
Neumann boundary conditions while $W$ the Dirichlet conditions.
The dual SYM operator is of the form
\beq{13}
 Q\Omega(Z'\Omega W\Omega)^{J}Q+\cdots.
\eeq

Moreover, the characteristic frequency of the fluctuations around
the classical solution~(\ref{13}),
$$\omega_n^2=n^2+2\nu^2+2m^2\pm 2\sqrt{(\nu^2+m^2)^2+n^2(\nu^2+2m^2)},$$
leads to spectrum of fluctuations as follows
  \beq{13a}
  \Delta E_o =\frac{\lambda}{8L^2}n\sqrt{n^2-4m^2}=\Delta E_c(2J_1,2J_2).
  \eeq

Again, in the SYM dual, we can use an integrable $SU(3)$ open spin
chain to calculate the anomalous dimension of the operators
(\ref{13}). Relative to the pseudo-vacuum that we have taken
before, there are $L=2J$ $\mu_{1,j}$-impurities and $J$
$\mu_{2,j}$-impurities in the ABAE. Then in the thermodynamic
limit, there should be two density functions for the Bethe roots,
one for each class of impurities \cite{YY69}, which will leads to
two coupled integral equations. To avoid this complication, we
note that the operators (\ref{13}) consist of $Z'$ and $W$ only,
so one can restrict to an $SU(2)$ open spin chain, with the
boundary terms $\Sigma_1=\Sigma(\otimes I_{2\times 2})^{L-1},\;
\Sigma_L=(I_{2\times 2}\otimes)^{L-1}\Sigma$, $\Sigma={\rm
diag}\{0,1\}$. Thus, we are allowed to take the pseudo-vacuum to
correspond to the operator consisting of $Z'$ only. The above
boundary terms are obtained by taking $n=2,\;l=1$ and
$\xi_+=0,\;\xi_-=1$ in the Hamiltonian~(\ref{24}). Then the
ABAE~(\ref{28}) reads
 \beq{33}
 \left(\frac{\mu_j+i}{\mu_j-i}\right)^{2(L+1)}=
 \prod^{J}_{\stackrel{k=1}{k\neq
 j}}\frac{\mu_j-\mu_k+2i}{\mu_j-\mu_k-2i}\frac{\mu_j+\mu_k+2i}
 {\mu_j+\mu_k-2i}.
 \eeq

It has been noted that the lowest energy eigenstates (with a fixed
impurity number) of the Hamiltonian~(\ref{24}) with a diagonal
$K$-matrix are also eigenstates of the shift operator
$\hat{t}=\hat{P}_{1,2}\hat{P}_{2,3}\cdots\hat{P}_{L-1,L}$~\cite{VG94},
where $\hat{P}$ is the permutation operator. Moreover, it is easy
to check that the operator~(\ref{13}) is unchanged if $\hat{t}$
acts on it twice. This observation yields, in the special case
with $L=2J$, the constraint
 \beq{33a}
 \prod^{J}_{j=1}\frac{\mu_j+i}{\mu_j-i}=\pm 1.
 \eeq
This constraint looks similar to the trace condition for a closed
chain, and implies the Bethe roots distribute symmetrically about
the origin.

The same way as the closed string/closed chain correspondence, we
expect that the imaginary Bethe roots in an open chain correspond
to a circular open string state. For simplicity, we consider the
case with $J$ odd. We may rescale the Bethe roots by
$\mu_j=2iq_jL$ with $q_j$ real, and take logarithm of the above
equations. For large $L$ they become
\beq{34}\frac{1}{q_j}=\frac{2}{L}
\sum^{\frac{J-1}{2}}_{\stackrel{k=-\frac{J-1}{2}} {k\neq j}}
\frac{1}{q_j-q_k}\quad\stackrel{k/J\to x}{\llraw}\quad\bint
dq\frac{\sigma(q)}{q_j-q}, \eeq where the integral is understood
as the principal value, and $\sigma(q)$ is the root density.

The important observation in \cite{BMSZ03} is that the imaginary
Bethe roots $\mu_j$ for a closed spinning chain form a
``condensate'' in the interior of an interval of order of $L$
along the imaginary axis around the origin. For the scaled
variable $q_j$, the condensate region is of order of unity.
Outside the condensate region, the Bethe roots start to spread
out. We expect a similar situation happens to the open chain as
well. Hence we make the following ansatz for the root density:
\beq{36}\sigma(q)=\left\{\begin{array}{ll} 2m
\quad\quad\quad\quad &-s<q<s,\;m=1,2,..., \\ \td{\sigma}(q) &s<q,
\\  \td{\sigma}(-q) &q<-s.
 \end{array}\right.
\eeq
Note our ansatz here is more general than that in \cite{BMSZ03}.
There are several remarks related to this ansatz:
\begin{itemize}
\item From the root density, it is easy to see that in the
condensed region, one has  $-i\mu_j\simeq 2j/m$, with integer
$j\in (-msL,\;msL)$. \item Part of equations in the
ABAE~(\ref{33}) can be reduced to
\beq{37}
\left(\frac{2j+m}{2j-m}\right)^{(L+1)}=
\prod^{msL}_{\stackrel{k=-msL}{k\neq
 j}}\frac{j-k+m}{j-k-m}\times [1+O(\frac{1}{L})].
 \eeq
For finite $j$ the left hand side of the above equations, in the
limit $L\to \infty$ increases (or decreases) exponentially; it has
to be compensated by a small denominator on the right hand side.
Therefore, $m$ has to be an integer. \item The $m=2$ case is the
same as for a closed string (or a closed chain) considered
in~\cite{BMSZ03}. There the roots $\mu\to\pm i$ have to be
included in order to ensure the trace condition if the root
$\mu_J=0$ is introduced. It implies that $m$ has to be
even$^3$\footnotetext[3]{Note that it is $m/2$ that is the winding
number of the closed circular string.}. \item It is
straightforward to check that the constraint~(\ref{33a}) is
satisfied with this ansatz, for both of $m$ even and $m$ odd. The
crucial point is that $\pm i$ are not roots for $m$ odd, then
$\mu_J=0$ contributes a factor of $-1$ to the left hand side of
(\ref{33a}).
\end{itemize}

With the ansatz~(\ref{36}) we can directly follow the method used
in sect. 4.1 of \cite{BMSZ03}, and obtain the solution of the
integral equation~(\ref{34}):
\beq{38}\td{\sigma}(q)=\frac{1}{\pi}\sqrt{q^2-s^2}
\left(-\frac{1}{qs}
+2m\int_{-s}^s\frac{dv}{(q-v)\sqrt{s^2-v^2}}\right), \eeq with the
consistency condition
\beq{39}\int_{-s}^s\frac{dv}{\sqrt{s^2-v^2}}=\frac{1}{2ms}. \eeq
The above consistency condition yields $s=1/2m\pi$. Substituting
Eq. (\ref{39}) into (\ref{38}), one can directly check that the
solution satisfies the normalization condition $\int
dq\sigma(q)=1$. The anomalous dimension is found, by using
(\ref{27}), (\ref{36}) and (\ref{38}), to be
\beq{40}\gamma&=&-\frac{\lambda}{8\pi^2}\left(\sum_{j=-msL}^{msL}
\frac{4m^2}{(2j)^2-m^2}+\frac{1}{L}\int_s^\infty
dv\frac{\td{\sigma}(v)}{v^2}\right) \nonumber \\
&=&\frac{m^2\lambda}{8L}. \eeq This result precisely matches the
spectrum for a circular solution~(\ref{10}) when $\lambda/L\ll 1$.
As mentioned above, $m$ could be any integer, even or odd, in the
open chain case.

Another way to solve this $(L,J)$ open chain system is to compare
the ABAE~(\ref{33}) with the ones for a $(2(L+1),2J)$ closed chain
system with Bethe roots distributing symmetrically about $\mu=0$.
In \cite{BMSZ03}, the state with an odd number of Bethe roots has
been studied thoroughly. One expects that the thermodynamic limit
is independent of whether the number of Bethe roots is odd or even
and, therefore, $E_o=\frac{1}{2}E_c(2L, 2J)$ holds true generally.

Finally let us consider the fluctuations around the ground
state~(\ref{40}). It corresponds to adding a few new roots or move
a few roots away from the imaginary axis. One lesson from the spin
chain description of the BMN spectrum is that the fluctuations
correspond to a set of Bethe roots with a few of them moving onto
the real axis. The spinless fluctuations, that is to move a few
roots from the imaginary axis to the real axis, do not change the
filling factor $\alpha=J/L=1/2$. To maintain the
constraint~(\ref{33a}), the roots should distribute symmetrically
about the origin. In other words, the roots are moved in pairs and
located on the real axis symmetrically. Let us consider a pair of
roots at $\pm 2\mu L$ with $\mu$ real and positive. In the large
$L$ limit, $\mu$ satisfies the equations
 \beq{42}\frac{1}{\mu}=n\pi+\int
dq\frac{\sigma(q)}{\mu-iq}. \eeq Substituting the solution
(\ref{38}) for the root density into the above equation, we
determine \beq{42a}\mu^{-1}=\pi\sqrt{n(n+4m)}, \eeq which
contributes to the anomalous dimensions
 \beq{43}\gamma_\mu=\frac{2\lambda}{8\pi^2
L^2\mu^2}= \frac{n(n+4m)\lambda}{4L^2}. \eeq On the other hand,
the roots at $\pm 2\mu L$ also back react on the roots on the
imaginary axis . It shifts $s$ to \beq{44}s\simeq
\frac{1}{2m\pi}\left(1-\frac{4}{L}
\frac{1}{\sqrt{1+(2m\pi\mu)^2}}\right), \eeq and the anomalous
dimension to \beq{45}\gamma_{\rm ir}=\frac{\lambda}{32\pi^2
L}\left(\frac{1}{s^2}
 -\frac{8}{L\mu^2}\left(1-\frac{1}{\sqrt{1+(2m\pi\mu)^2}}\right)\right).
\eeq Substituting Eqs.~(\ref{42a}) and (\ref{44}) into (\ref{45})
and adding (\ref{43}), we find that the change in the anomalous
dimension~(\ref{40}) is
\beq{46}\Delta\gamma=\frac{\lambda}{4L^2}(n+2m)\sqrt{n(n+4m)}
=\frac{\lambda}{4L^2}n'\sqrt{n'^2-4m^2}. \eeq This precisely
doubles the spectrum~(\ref{13a}) predicted by the fluctuations of
open circular string. It just reflects the fact that two
fluctuation modes have been counted.

If we add a few new roots on the real axis, the total filling
factor$^5$\footnotetext[5]{The change in the filling factor of
order $1/L$ will change the ground state energy by an amount of
order $1/L^3$ only\cite{BMSZ03}.} will become greater than $1/2$.
Therefore, the constraint~(\ref{33a}) will be broken down. In
other words, the new roots at the real axis need not to be in
pairs and be symmetrical. The simplest case is to consider a
single new root added. The calculation, however, is much more
subtle than that for spinless fluctuations. We do not present the
details here, but merely mention we have verified that the change
in the anomalous dimension~(\ref{40}) is indeed half of the one in
(\ref{46}), as expected.

\section{Conclusion}

In \cite{CWW03}, we have identified an integrable structure in a
conformal ${\CN 2}$ $Sp(N/2)$ gauge theory, at least at planar
one-loop level. In the open-string holomorphic scalar sector, it
is an integrable $SU(3)$ open spin chain. In this letter, we
explore this structure to test the AdS/CFT duality for open
spinning strings. It is shown that the spectrum of both a circular
and folded open spinning string can be reproduced at planar
one-loop level in SYM by examining the thermodynamic limit of the
Bethe ansatz for the open spin chain, in the same spirit as in
ref. \cite{YY69}. We have verified that the classical energy of
the open spinning string is always related to that of a closed
string counterpart by
 \beq{41}
 E_o=\frac{1}{2}E_c(2J_1,2J_2, \cdots).
 \eeq
This relation was first found in three-spin solutions in
\cite{Stef03b}. Moreover, we also verified that fluctuation
spectrum around classical circular open string states can be
reproduced in SYM by means of the open spin chain. The fluctuation
energy of open spinning string is related to that of a closed
string counterpart by
 \beq{41}
 \Delta E_o=\Delta E_c(2J_1,2J_2, \cdots).
 \eeq
It is slight different from the relation between classical energy.
From the open spin chain's point of view, both of two relations
reflects the doubling trick, namely the length of the chain gets
doubled and the impurities form mirror images, due to the
existence of the ends of the chain. From the viewpoint of string
theory, it just reflects the doubling trick in 2d conformal field
theory with boundary. Roughly speaking, a closed string can be
constructed from two copies of an open string. Thus, the validity
of this relation in both approaches naturally suggests a more
direct relation between a string and a spin chain; i.e. a spin
chain may be viewed as a discrete version of a string. More
examples to check this relation are welcome.

There exist three-spin open spinning solutions, as shown in
\cite{Stef03b}. It would be very interesting to study the SYM
duals of such solutions in terms of the open spin chain.

\acknowledgments{ Y.-S.~Wu thanks the Interdisciplinary Center for
Theoretical Study, Chinese Academy of Sciences for warm
hospitality and support during his visit, when the collaboration
began to form. This work is partly supported by the NSF of China,
Grant No. 10305017, and through USTC ICTS by grants from the
Chinese Academy of Science and a grant from NSFC of China.}

\end{document}